\documentclass[doublecol]{epl2}

\def\real{I\!\!R}

\title {A ``Gaussian'' for diffusion on the sphere}
\shorttitle{A ``Gaussian'' for Diffusion on the sphere
} 

\author{Abhijit Ghosh
$^{1}$,  Joseph Samuel$^{2}$\and
Supurna Sinha$^{2}$}
\shortauthor{Abhijit Ghosh et al}

\institute{\inst{1}
Department of Chemical and Biomolecular Engineering
Sogang University, \\ \,\,\,\, 35, Baekbeom-ro, Singsu-dong, Mapo-gu, 
Seoul, South 
Korea\\
\inst{2} Raman Research Institute, C. V. Raman Avenue, Sadashivanagar, 
Bangalore, India 560 080.}


\abstract{We present an analytical closed form expression,
which gives a good approximate propagator for diffusion on the sphere. 
Our formula is the spherical counterpart of the Gaussian propagator for
diffusion on the plane. While the analytical formula is derived
using saddle point methods for short times, it works well even for 
intermediate times. Our formula goes beyond 
conventional ``short time heat kernel expansions" in that it is 
nonperturbative
in the spatial coordinate, a feature that is ideal for studying large 
deviations. Our work suggests a new and efficient algorithm for 
numerical integration of the diffusion equation on a sphere. 
We perform Monte Carlo simulations to compare the numerical efficiency 
of the new algorithm with the older Gaussian one. 
} 
\begin{document}

\maketitle

{\bf Introduction:} Diffusion on the sphere is a problem that arises
in several contexts. At the cellular level, diffusion is an 
important
mode of transport of substances. The cell wall
is a lipid membrane and biological substances like lipids and proteins
diffuse on it. 
In general biological membranes are curved 
surfaces. Spherical diffusion
also crops up in the swimming of bacteria, surface smoothening in 
computer graphics\cite{bulow} and
global migration patterns of marine mammals\cite{brill}. 
Such diffusion effects are studied using computer simulations.  
There have been experimental studies of 
fluorescent marker molecules on curved surfaces like micelles and 
vesicles\cite{mmgk1,mmgk2} using fluorescence anisotropy decay.

While diffusion in Euclidean space has been studied extensively 
both analytically and numerically\cite{wax, tobochnik},
there have been fewer studies of diffusion on curved surfaces.
There have been some analytical studies of diffusion on curved surfaces
\cite{jb,faraudo,bm}. 
Some of these studies\cite{jb,faraudo} 
arrive at a short time  heat kernel expansion 
\cite{risken,birrell,drozdov} for 
the probability distribution. This expansion is perturbative in the time step
as well as the spatial step as is evident from Eq. (51)
of Ref.\cite{faraudo}.
However, there does not exist a simple 
closed form analytical expression, for the probability distribution for diffusion on a curved 
surface which can be implemented in a computer simulation.
Here we focus on diffusion on a sphere, the case of 
a curved surface with constant positive curvature.     
The simplest and most natural method of generating diffusion on a curved 
manifold is to consider a large number of walkers 
performing a random motion. 
The random walkers can be described by 
a Langevin equation using a spatial step size much smaller than the inverse 
curvature so that curvature can be neglected. 
In this ``tangent space''
approximation, the sphere appears locally flat and one can ignore its
curvature. This is the approach taken for example in 
\cite{mmgk1,mmgk2} in which the diffusing particle takes a Brownian 
step in the tangent plane at a point on $S^2$ and is then projected back 
to the surface of the sphere.
One can describe the ensemble of random walkers with a probability 
distribution
$P$ and write a Fokker-Planck (diffusion) equation governing the spread 
of probability. One finds (see page 847 of Ref. \cite{nrc})
on general grounds, the temporal step size must be 
chosen smaller than the square of the spatial
step size. As a result of the small spatial step size forced 
on us by the tangent space method, integrating in time is 
computationally expensive. It would be useful to have a method
which permits larger step sizes, not constrained by the curvature.

There has been some work \cite{nissfolk, carlsson} generating an  
algorithm for diffusion on $S^2$ by making use of known analytic results
on $S^3$, the hypersphere in four dimensions and 
identifying $S^2$ with a thin strip around the equator of $S^3$ with   
a reflecting boundary. These methods give rise to an improved algorithm
for treating diffusion on $S^2$. However, the use of $S^3$ and the 
method of dimensional reduction render the treatment less than 
intuitive. Our purpose here is to provide a more down to 
earth approach by working {\it directly} on $S^2$. We 
derive an approximate
analytical formula for the distribution function representing diffusion on
a sphere. 
In contrast to earlier studies\cite{jb,faraudo} we use a 
semiclassical saddle point approximation which is perturbative in time
but {\it nonperturbative in space}. Our formula reduces to the 
formula presented in \cite{jb,faraudo} in the limit of small spatial 
scales.
In the planar limit our formula reduces to a Gaussian. 
We use our ``Gaussian'' on the sphere to generate an efficient 
algorithm 
for simulating Brownian motion on a sphere. The algorithm improves over 
the earlier tangent space simulation algorithm.
We expect this algorithm to have a wide range of applications,
especially in fields where one needs to repeatedly generate diffusion  
processes on spheres. 
In the physics of 
stiff, inextensible polymers, one  
has contraints on the bond lengths while the bond angles are relatively 
unconstrained. This leads to random motions, as in the Kratky-Porod 
model which represent diffusion on  spheres. In computer graphics,
one seeks algorithms which  can be used to smooth graphical data 
on spherical images. A standard technique is to use the diffusion
equation on the sphere, but a simple and accurate algorithm for 
implementing this is presently lacking.

{\bf Theory:} To define our problem regarding diffusion on 
a sphere, consider first
diffusion on a plane which is described by
\begin{equation}
\frac{\partial P}{\partial t} (\vec{x}, t) = D\nabla^{2} P (\vec{x}, t)
\label{plan}
\end{equation}
where $D$ is the diffusion constant and \(P (\vec{x}, t)\) the probability
distribution on the plane at time $t$. The elementary solution
with initial distribution \(P (\vec{x}, 0) = \delta^{2} (\vec{x}) \) is 
easily found by Fourier transform. The formal solution is 
\begin{equation}
P (\vec{x},t) = \frac{1}{4\pi^2}\int d\vec{k}\; e^{-{D\vec{k}\cdot 
\vec{k}t}}\; e^{i\vec{k}\cdot \vec{x}}
\label{fourier}
\end{equation}
which can be integrated to give a simple Gaussian form
\begin{equation}
P (\vec{x},t) = \left(\frac{1}{4\pi D t}\right)\; \exp 
-\frac{\vec{x}\cdot 
\vec{x}}{4 D t} 
\label{gaussianplanarcart}
\end{equation}
When expressed in plane polar co-ordinates $(r,\theta)$, the probability 
distribution function in $r$ is
\begin{equation}
P (r,t) = \left(\frac{r}{2 D t}\right)\; \exp 
-\frac{r^2}{4 D t} 
\label{gaussianplanarpolar}
\end{equation}
While these two expressions (\ref{fourier},\ref{gaussianplanarpolar}) 
are 
formally equal to 
each other, the first is an unwieldy integral whereas the second is a 
simple closed form 
Gaussian expression which is well
suited to numerical work. In
(\ref{fourier}) the time $t$ appears in the 
numerator of the exponent leading to slow convergence for small times,
unlike (\ref{gaussianplanarpolar}) where it appears in
the denominator.
Our purpose in this Letter is to seek the 
analogue of (\ref{gaussianplanarpolar}) for spherical diffusion.

Consider the analogous situation on the sphere 
\begin{equation}
{\cal S}^{2} = \left\{(x^{1}, x^{2}, x^{3})\; \epsilon \real^{3} : 
(x^{1})^{2} + 
(x^{2})^{2} + (x^{3})^{2} = R^{2} \right\}
\end{equation}
of radius $R$ in three dimensional Euclidean space. In standard polar 
coordinates 
on the sphere the diffusion equation is
\begin{equation}
\frac{\partial P}{\partial t} = \frac{D}{R^{2}}
 \;\left( \frac{1}{\sin\theta} \frac{\partial} {\partial\theta} 
\sin\theta 
\frac{\partial P}{\partial\theta}
+ \frac{1}{\sin^{2}\theta}\; \frac{{\partial}^{2} 
P}{\partial\varphi^{2}}\right)
\end{equation}\\
Let us define a dimensionless time
variable $\tau = 2D t/R^2$ for convenience. 
Typical values for diffusion on a lipid membrane are
$R=1\mu{\rm m}$ and $D=1\mu{\rm m}^2$/sec. The $\tau$ values
of interest to us are on the order of seconds.

For an initial $\theta$ distribution $P(\theta,0)$  
(including the measure $\sin{\theta}$) with $\delta$ 
function support at the North pole $\theta=0$, the diffusion equation 
has a  formal solution for the final distribution in $\theta$ 
at time $\tau$:  

\begin{equation}
P(\theta,\tau) =\frac{\sin{\theta}}{2}
\sum_{l=0}^{\infty}(2l+1)e^{-l(l+1) 
\tau/2}P_l(\cos{\theta}).
\label{exact}
\end{equation}
Such a solution is formally exact and analogous to 
(\ref{fourier}), but
in practice, unwieldy to use in numerical work since it is 
expressed
as an infinite series. For short times, the convergence of the series
is poor and truncation leads to spurious oscillations. 

Is there an analogue of the Gaussian form (\ref{gaussianplanarpolar})
on the sphere? The main result of this Letter is a closed form 
approximate expression
which generalises the planar Gaussian (\ref{gaussianplanarpolar}) to the 
sphere. 
This approximate expression is given by:
\begin{equation}
Q(\theta,\tau) ={\frac{\cal 
N(\tau)}{\tau}}\sqrt{\theta\sin{\theta}}e^{-{\frac{{\theta}^2}{2\tau}}}
\label{approx}
\end{equation}
where ${\cal N}(\tau)$ is a constant determined by the normalisation
condition
\begin{equation}
\int_0^\pi d\theta Q(\theta,\tau)=1
\label{normalisation}
\end{equation}
In the limit of small times, the expression (\ref{approx}) reduces to 
$\frac{1}{\tau} \theta \exp{-\frac{\theta^2}{2\tau}}$
which agrees with the planar 
Gaussian (\ref{gaussianplanarpolar}). Our study shows that this 
expression 
turns out to be a good approximation to the exact propagator 
(\ref{exact}) even for intermediate times.

The Heat Kernel Expansion (HKE) results in a propagator (see 
Eq. 51 in 
Ref.[\cite{faraudo}]) expressed as a power series in $\theta$
\begin{equation}
Q_{HKE}(\theta,\tau) 
=\frac{1}{\tau} 
(1+\frac{\theta^2}{12}+\dots) 
e^{-{\frac{{\theta}^2}{2\tau}}}\sin{\theta}
\label{HKE}
\end{equation}
where  we have transcribed the content of 
Eq. 51 in Ref.[\cite{faraudo}] (suitably restricted to constant 
curvature) and  inserted the measure factor of $\sin{\theta}$ for easy 
comparison. The new propagator is essentially a summation
of this perturbative series.

\begin{figure}
\onefigure[scale=0.4]{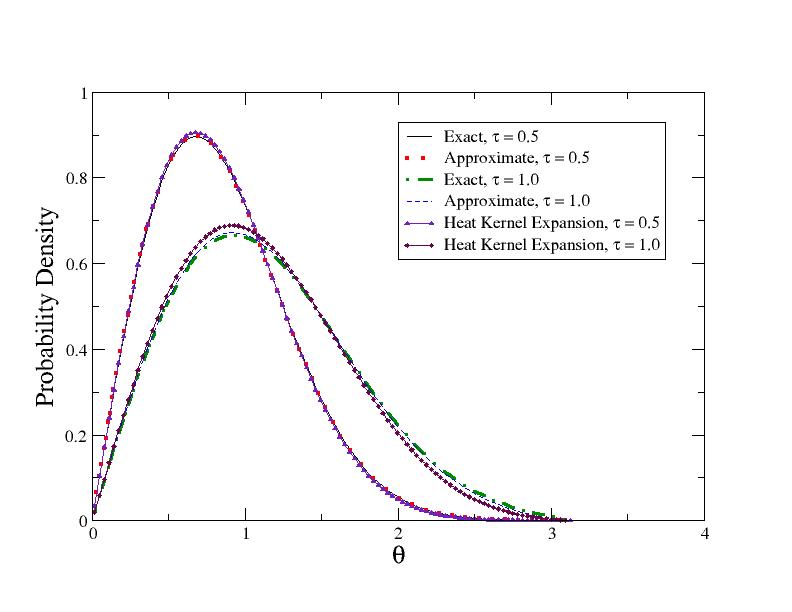}
\caption{(Colour online) 
Figure shows a comparison of the exact
propagator (Eq. (\ref{exact}) truncated to $l_{max}=20$), the 
approximate one Eq. (\ref{approx}) and the Heat Kernel expansion for $\tau=0.5,1.0$. 
Note that the difference between the first two is barely discernible.
}
\label{theocomparison1}
\end{figure}

Figures 1 and 2 show a 
comparison of the exact propagator (with the upper limit $l_{max}$ in 
the summation set to $20$) with the approximate
one for a range of $\tau$.
The plots of Fig. (\ref{theocomparison1}) show $P(\theta,\tau)$ and 
$Q(\theta,\tau)$  for $\tau=.5,1$ and $Q(\theta,\tau)_{HKE}$ truncated 
to the second term. 

The deviation between 
probability 
distributions can be quantitatively measured using the Kullback-Leibler 
divergence,
which is also known as the relative entropy. This measure which is
widely used in information theory \cite{cover} gives a positive number
$D_{KL}$ which measures the extent of deviation (or divergence) between 
a trial distribution $Q(\theta)$ and a fiducial one $P(\theta)$:
\begin{equation}
D_{KL}:= \int_0^\pi d\theta P(\theta) 
\log{[\frac{P(\theta)}{Q(\theta)}}]
\label{Kullback}
\end{equation} 
$D_{KL}$ vanishes if 
and only
if the two normalised distributions are identical. $D_{KL}$ is also 
invariant
under changes of coordinates. 
For the fiducial distribution we use the exact form 
(\ref{exact}) and   
compare KL divergence of 
the Gaussian propagator (restricted 
to the interval $[0,\pi]$ and normalised in this interval), 
the $Q_{HKE}$ (\ref{HKE}) as well as our new distribution $Q_{new}$.
The table shows a comparison of $D_{KL}$ Gaussian, 
$D_{KL}$ HKE  and $D_{KL}$ new,
 for different values of 
$\tau$.

Note that the new propagator always has a smaller KL divergence
than the Gaussian propagator and $Q_{HKE}$ . 
It is thus a better approximation to
the exact propagator. 
\begin{center}
\begin{tabular}{|l|l|c|r|}
      \hline
      time $\tau$ & $D_{KL}$ Gaussian & $D_{KL}$ HKE & $D_{KL}$ new\\
       \hline      
       0.1 & 1.4$\times$ 10$^{-4}$ & 6.5 $\times$ 10$^{-7}$ & 1.8 $\times$ 
10$^{-9}$\\
      0.5 & 4.0$\times$ 10$^{-3}$ & 3.7 $\times$ 10$^{-4}$ & 3.8 $\times$ 
10$^{-6}$\\
      0.7 & 8.2$\times$ 10$^{-3}$ & 1.3$\times$ 10$^{-3}$ & 2.9$\times$ 
10$^{-5}$\\
      0.9 & 1.3$\times$ 10$^{-2}$& 2.9$\times$ 10$^{-3}$ & 9.9$\times$ 
10$^{-5}$\\
      1.0 & 1.6$\times$ 10$^{-2}$& 3.9 $\times$ 10$^{-3}$ & 1.6 $\times$ 
10$^{-4}$\\
      2.0 & 4.3$\times$ 10$^{-2}$& 1.1$\times$ 10$^{-2}$ & 8.9$\times$ 
10$^{-4}$\\
      \hline
    \end{tabular}
\end{center}

{\bf Derivation of the Main Result:}
The elementary solution to the diffusion equation on the unit sphere is 
given
by the 
Wiener integral\cite{kleinert}
\begin{equation}
K(\hat{n}_{1},\;0,\;\hat{n}_{2},\;\tau) = \int {\cal D} [\hat{n} (\tau)] 
\exp - S [\hat{n} (\tau)]
\end{equation}
where  
\begin{equation}
S [\hat{n}(\tau)] = 
\frac{1}{2} \int^{\tau}_{0} \frac{d \hat{n}}{d\tau}.\frac{d 
\hat{n}}{d\tau} d\tau.
\end{equation}
$K(\hat{n}_{1},\;0,\;\hat{n}_{2},\;\tau)$ is the conditional 
probability that the particle will be at $\hat{n}_{2}$ at time
$\tau$ given that it was at $\hat{n}_{1}$ at time 0. 
For short times $\tau$ , we may use
a semiclassical approximation
\cite{schulman}
\begin{equation}
K(\hat{n}_{1},\;0,\;\hat{n}_{2},\;\tau) ={\cal N(\tau)} 
\sqrt{\det V}\; \exp 
- S_{\rm cl} [\hat{n}_{1}, \hat{n}_{2}, \tau]
\end{equation}
where \(S_{\rm cl} [\hat{n}_{1}, \hat{n}_{2}, \tau)]\) is defined as the 
classical 
action
of the least action path connecting \(\hat{n}_{1}\) to \(\hat{n}_{2}\) in
``time" \(\tau\). The least action path is unique if $\hat{n}_{1}, 
\hat{n}_{2}$ are 
not collinear, which we assume and thus exclude the isolated points $ 
\theta=0,\pi$.
Det \(V\) is the Van Vleck determinant given by the determinant of
the $2\times2$ Hessian matrix
\begin{equation}
V_{ij} = \frac{\partial^{2} S_{\rm cl} [\hat{n}_{1},\; \hat{n}_{2},\; 
\tau]}{\partial\hat{n}^{i}_{1}\; \partial\hat{n}^{j}_{2}}
\end{equation}
and ${\cal N}$ is a normalisation constant.

Varying the action to find the classical path yields the geodesic
equation
\begin{equation}
\frac{d^2 \hat{n}}{d \tau^2} = \lambda\hat{n}, \label{geo}
\end{equation}
where $\lambda$ is a Lagrange multiplier enforcing the constraint 
\(\hat{n} \cdot \hat{n} = 1. \)
The solution to (\ref{geo}) is the unique great circle passing through 
$\hat{n}_{1}$ and
$\hat{n}_{2}$. The classical action is given by 
\begin{equation}
S_{\rm cl} = \theta^{2}/2\tau
\end{equation}
where \( \cos\theta = \hat{n}_{1}\cdot \hat{n}_{2}\) defines  $\theta$,
\( 0 < \theta < \pi\), the length of the shortest geodesic arc 
connecting 
\(\hat{n}_{1}\) to \(\hat{n}_{2}\). 
From an evaluation of the Van Vleck determinant 
\begin{eqnarray*}
\det V
&=& \hat{n}_{1p}\;\epsilon^{pil}\;\hat{n}_{2q} 
\;\epsilon^{qjk}\;V_{ij}\;V_{lk}
\end{eqnarray*}
we find that 
\begin{equation}
\det V = \frac{\theta}{\tau^{2}\sin\theta}.
\end{equation}
This leads to the approximate propagator
\begin{equation}
K[\hat{n}_{1}, 0, \hat{n}_{2}, \tau] = \frac{{\cal 
N(\tau)}}{\tau}\; 
\sqrt{\theta/\sin\theta}\;\; e^{-\frac{\theta^{2}}{2\tau}}.
\label{approxprop}
\end{equation}
Multiplying by the measure $\sin{\theta}$ to convert into a $\theta$
distribution leads to the approximate propagator
which is used in eq.(\ref{approx}).


{\bf Computer Simulations:}
We have performed Monte Carlo simulations of spherical diffusion to 
investigate the numerical efficacy of the approximate propagator derived 
analytically in the last section. 
The Fokker-Planck 
equation on a sphere is exactly solvable and 
the solution can be expressed as
a series involving the Legendre polynomials as described 
earlier. However, 
because of poor convergence of the series for short times, 
this form is not suitable for repeated numerical 
evaluation. 
We use the exact solution with a comfortably large cutoff
$l_{max}=20$ as a standard against which the 
computer simulations are tested. 

In performing the simulations, we have closely 
followed the method adopted in Ref.\cite{mmgk1,mmgk2} for 
studying diffusion of fluorescent molecules on a 
sphere. We have studied the diffusion of $5\times10^6$ molecules and 
obtained the distribution of the final polar angular displacement 
$\theta$. 

To integrate for a time 
$\tau$, we split up the time interval $\tau$ into 
$n_{step}$ time 
steps $\tau=n_{step}\tau_{step}$ each of size $\tau_{step}$. 
The value of $\tau_{step}$ determines the spatial step size $\sigma$ via 
the standard diffusion relation
\begin{equation}
\tau_{step}=\frac{\sigma^2}{2}
\label{taurelation}
\end{equation} 
We have implemented 
two distinct algorithms for randomly choosing $\beta$: a Gaussian 
algorithm and our proposed new algorithm. 
In the first algorithm, we replace the spherical geometry locally by the 
tangent plane to the sphere at 
the starting point of each Monte Carlo move. 
Confining ourselves to  the local tangent plane 
on the sphere at each move, 
we choose $\beta$ according to  the distribution 
(eq. \ref{gaussianplanarpolar})
\begin{equation}  
Q_{Gauss}({\beta})=\frac{2\beta}{\sigma^2} \exp(-\beta^2/\sigma^2).
\label{gaussalpha}
\end{equation} 
Thus at each step, the angular displacement $\beta$ is chosen 
from a two dimensional Gaussian distribution which has zero mean and 
standard deviation $\sigma$. This is our first  algorithm. It is only
expected to be accurate when the step size $\sigma$ is small.

In the second method, the angular displacement at each step was chosen 
from the approximate propagator given by (\ref{approx}).
\begin{equation}
Q_{new}({\beta})=\frac{2 {\cal N}}{\sigma^2} \sqrt{\beta\sin\beta} 
\exp{(-\beta^2/\sigma^2)}
\label{approxalpha}
\end{equation} 
This distribution reduces to the previous one (\ref{gaussalpha}) when 
the step size $\sigma$ is small, but has a larger range of validity
since it takes into account the curvature of the sphere. 

In both algorithms the random numbers were generated by 
constructing a random walk of the variable  $\beta$ in the external 
potential $-log Q(\beta)$. After discarding $10^5$ steps to get rid of 
transients, we arrange the subsequent  
$10^6$ $\beta$ values as a $10^3\times10^3$ matrix, transpose it 
and store the values for use in the program. This shuffles the 
random numbers and removes correlations between neighboring steps. 

Whereas the planar Gaussian algorithm needs to use small displacements 
to ensure that the tangent plane approximation remains valid, 
the proposed new algorithm  is not constrained by this requirement. It
remains a good approximation even for intermediate $\sigma$.
As a result in order to integrate for $\tau=2$, we need to use 
just two time steps with a step size of $\sigma=1.414$ with the new 
algorithm (Figure 3). With the Gaussian algorithm, 
even with eight time steps of step size $\sigma=.7071$, one achieves a 
poorer accuracy (Figure 4).
This new algorithm is our second main point in this Letter.

\section{Conclusion}
In this Letter we derive an approximate
analytical formula for the distribution function for spherical diffusion 
and use it to generate a simple and efficient algorithm for 
simulations. 
While we have restricted ourselves to spherical diffusion, it is evident
that our results apply equally well to the saddle, (or hyperbolic 
plane),
the space of constant negative curvature. The only change is that
the circular function $\sin{\theta}$ is replaced by the hyperbolic 
function $\sinh{\theta}$.  Our results for the approximate
propagator (\ref{approxprop}) also extend to higher 
(D) dimensional spheres and saddles
with the slight modification that the prefactor 
$\theta/\sin\theta$ (or its hyperbolic form) is replaced by 
$(\theta/\sin\theta)^{\frac{D-1}{2}}$.

\revision{Rasin et al\cite{rasin} studied diffusion on a planar 
lattice and addressed the problem of scaling of
the time step as the square of the lattice size. Their solution 
uses kinetic methods that replace the parabolic diffusion equation
by a hyperbolic one, thus allowing for larger time evolution 
steps. It would be interesting to extend their 
work, using a variable coordination number on the lattice to encode 
curvature\cite{bowick}.}  

For a general curved surface embedded in three dimensional space, 
the intrinsic geometry is determined entirely by the Gaussian curvature 
$\kappa$. 
Since diffusion on the surface depends {\it only} on the intrinsic 
geometry,
we may consider each part of the surface as locally 
approximated by a sphere of radius $R=1/\sqrt{\kappa}$. 
We may then apply the 
new simulation algorithm described in this paper in a local patch. This 
entails numerically introducing geodesic based coordinates and choosing
isotropically distributed steps, with a size distribution given by
(\ref{approxalpha}). 
We expect this to improve on the tangent space simulation methods,
since these only approximate the local geometry to first order 
and do not take into account the second order effects of curvature. 
The new propagator would be useful in a wide variety of 
applications.

\acknowledgments
It is a pleasure to acknowledge many stimulating discussions
with Abhishek Dhar, Rajaram Nityananda and Sanjib Sabhapandit.
\bibliographystyle{eplbib}
\bibliography{refs1}

\onecolumn

\begin{figure}
\onefigure[scale=0.4]{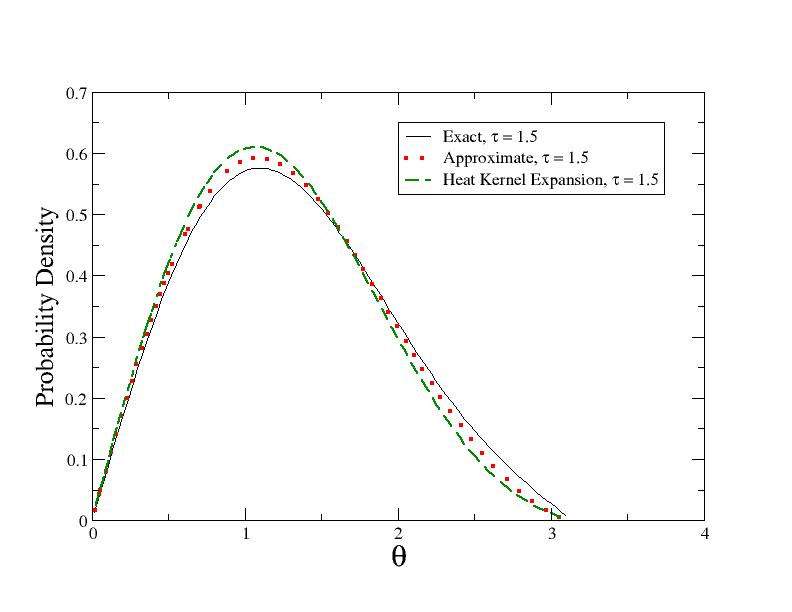}
\caption{(Colour online) Figure shows a comparison of the exact
propagator (Eq. (\ref{exact}) truncated to $l_{max}=20$) with the 
approximate one (Eq. 
(\ref{approx})) for $\tau=1.5$. The difference is now apparent.
Also shown is the Heat Kernel expansion, which differs even more from the exact propagator.
}
\label{theocomparison2}
\end{figure}

\begin{figure}
\onefigure[scale=0.4]{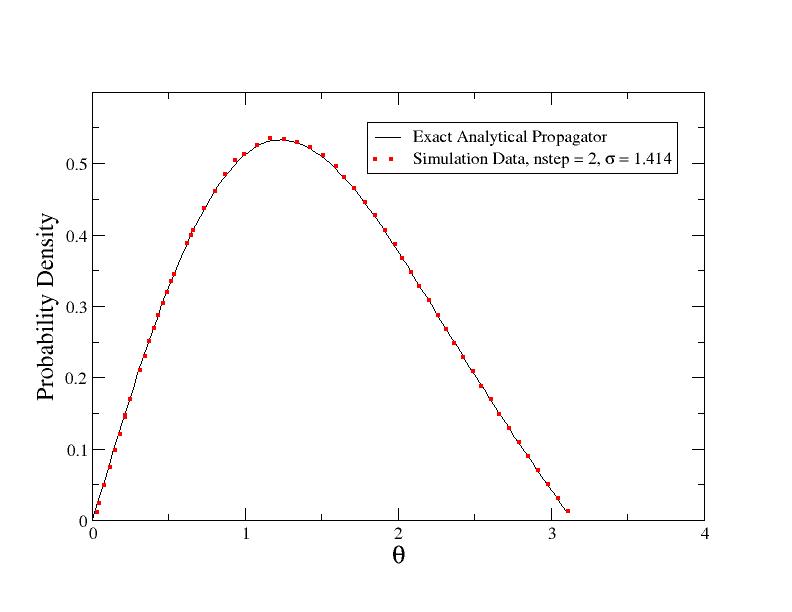}
\caption{(Colour online) 
Figure shows a comparison of the exact propagator with 
the simulation data using the new algorithm. The total time $\tau$
of integration is $\tau=2$ and this time has been achieved with
two steps, each  of step size $\sigma=1.414$.
} 
\label{twostepapprox}
\end{figure}
\begin{figure}
\onefigure[scale=0.4]{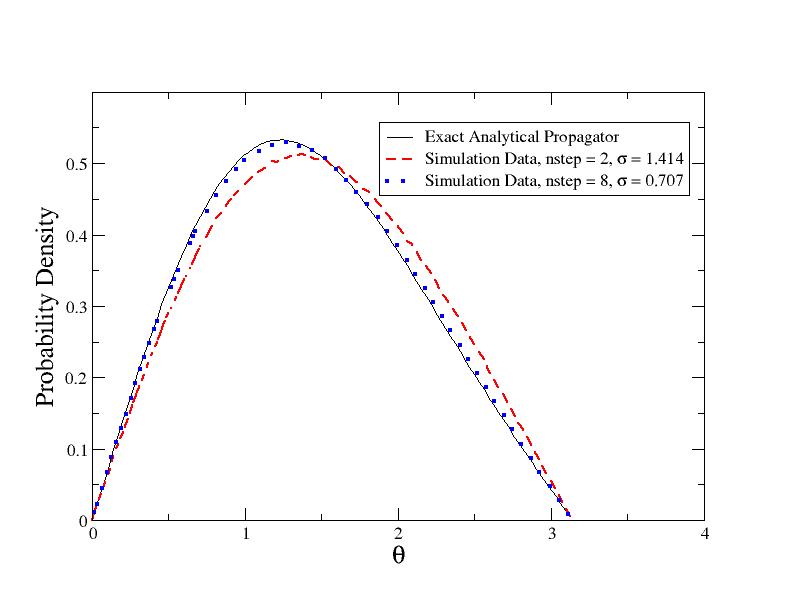}
\caption{(Colour online) 
Figure shows a comparison of the exact propagator with 
the simulation data using the Gaussian algorithm. The total time $\tau$
of integration is $\tau=2$ and this time has been achieved with
two steps of step size $\sigma=1.414$ (dashed curve) and eight steps
of size $\sigma=.7071$. Notice that the Gaussian algorithm requires a 
larger number of steps to reproduce the exact analytical propagator. 
} 
\label{2and8step with exact}
\end{figure}

\end{document}